# Effects of Fine Particulate Matter on Cardiovascular Disease Morbidity: A Study on Seven Metropolitan Cities in South Korea

Eunjung Cho[1], Yeonggyeong Kang[2] and Youngsang Cho[1]*

[1]Department of Industrial Engineering, Yonsei University, Seoul, South Korea, [2]Korean Foundation for Quality, Seoul, South Korea

**Objectives:** The primary purpose of this study is to analyze the relationship between the first occurrence of hospitalization for cardiovascular disease (CVD) and particulate matter less than 2.5 μm in diameter (PM2.5) exposure, considering average PM2.5 concentration and the frequency of high PM2.5 concentration simultaneously.

**Methods:** We used large-scale cohort data from seven metropolitan cities in South Korea. We estimated hazard ratios (HRs) and 95% confidence intervals (CIs) using the Cox proportional-hazards model, including annual average PM2.5 and annual hours of PM2.5 concentration exceeding 55.5 μg/m$^3$ (FH55).

**Results:** We found that the risk was elevated by 11.6% (95% CI, 9.7–13.6) for all CVD per 2.9 μg/m$^3$ increase of average PM2.5. In addition, a 94-h increase in FH55 increased the risk of all CVD by 3.8% (95% CI, 2.8–4.7). Regarding stroke, we found that people who were older and had a history of hypertension were more vulnerable to PM2.5 exposure.

**Conclusion:** Based on the findings, we conclude that accurate forecasting, information dissemination, and timely warning of high concentrations of PM2.5 at the national level may reduce the risk of CVD occurrence.

## INTRODUCTION

Interest in particulate matter (PM) is growing as it is one of the most lethal threats to human health. PM can penetrate deeper into the body through the respiratory tract as its size decreases (1). Fine particles of 1–5 μm in diameter can enter the body and spread to the blood vessels, causing cardiovascular disease (CVD) (2). Several studies have analyzed the relationship between average PM less than 2.5 μm in diameter (PM2.5) concentration and CVD mortality (3–6) and morbidity (7–10).

Owing to the clear evidence that PM has harmful health effects, the Korean government has attempted to reduce the levels of PM exposure for people. Nevertheless, according to a survey by the Korean Environmental Institute, most Koreans thought that the PM concentration in 2018 was higher than it was 10 years ago (11). However, the monitored annual average PM less than 10 μm in diameter (PM10) concentration in Korea declined from approximately 54 μg/m$^3$ in 2008 to approximately 41 μg/m$^3$ in 2018 (12). Similarly, the annual average PM2.5 concentration declined from approximately 26 μg/m$^3$ in 2015 to approximately 23 μg/m$^3$ in 2018. This discrepancy between the public perception of PM concentration and the monitored trend of annual average PM2.5 concentration appears to be due to the increased frequency of exposure to high PM concentrations in a short period of time. Currently, in Korea, the PM2.5 alert is issued when the hourly average PM2.5 concentration exceeds 75 μg/m$^3$ for more than 2 hours and is withdrawn when it drops below 35 μg/m$^3$. Prior to July 1, 2018, these criteria were 90 μg/m$^3$ and 50 μg/m$^3$, respectively. **Figure 1** depicts the trend of PM2.5 alerts in some major cities in Korea from 2016 to 2019 (12). The total days when the alert was issued and then withdrawn increased from 2016 to 2019. More details can be found in the **Supplementary Table S1**. In addition, the annual hours when the average PM2.5 concentration exceeded 55.5 μg/m$^3$ in Seoul, Korea increased from 4.1% in 2016 to 6.5% in 2019 (13).

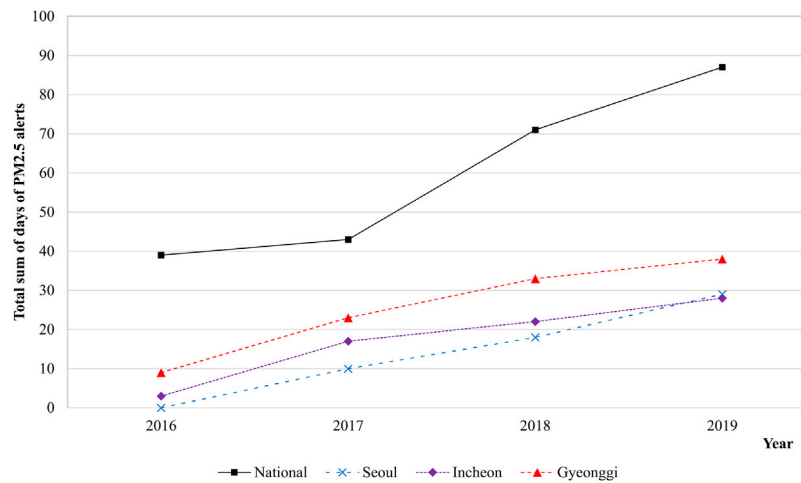

**FIGURE 1** | Particulate matter alerts trends by major cities in South Korea between 2016 and 2019 (Source: Ministry of Environment [12]). Abbreviation: PM2.5, particulate matter less than 2.5 μm in diameter. Alert issuance criteria: the hourly average PM2.5 concentration is above 90 μg/m$^3$ for over 2 h (Before July 2018); the hourly average PM2.5 concentration is above 75 μg/m$^3$ for over 2 h (Since July 2018).

Although the frequency of high PM2.5 concentration is increasing, to the best of our knowledge, no study has analyzed the relationship between PM2.5 exposure and CVD using the frequency of high PM2.5 concentration and average PM2.5 concentration simultaneously. Some studies used the sub-time PM concentration, that is, they used PM information for a specific time as a variable instead of annual average PM concentration; their findings have revealed that the sub-time variable is significantly related to human health (14–16). In contrast to these studies, in this study, we analyzed the relationship between the first occurrence of CVD hospitalization and the frequency of high PM2.5 concentration. We used the Cox proportional-hazards model using cohort data in Korea. To accurately analyze the effect of high PM2.5 concentration exposure, we used the annual exposure hours of high PM2.5 concentration and average PM2.5 concentration in the model, considering the individual residential district if the participant moved their residence. This study is structured as follows. We explain the data and models used in this study in **Section 2**, and provide the estimation results in **Section 3**. Finally, we discuss the implications, limitations, and future research directions in **Section 4**.

## METHODS

### Study Population, Health Outcomes, and Air Pollutant Data

We used cohort data provided by the National Health Insurance Service (NHIS) of Korea. All residents in Korea are obliged to partake in the NHIS; thus, NHIS cohort data covers the entire population living in Korea, including Koreans and foreigners. There are two types of cohort data provided by the NHIS: the NHIS-National Sample Cohort (NHIS-NSC) and NHIS-National Health Information Database (NHIS-NHID) (17, 18). The NHIS-NSC are randomly selected data from approximately one million people of the total population in 2002–2015; thus, the data are reduced and refined for the researcher's convenience. The NHIS-NHID contains data that can be extracted depending on specific patient characteristics according to the needs of the researcher. Therefore, it is useful for analyzing specific diseases during a specific period, although the data are raw and include missing values and errors. In this study, we used NHIS-NHID data because we targeted a specific disease—CVD—and a specific period, 2015–2018. We refined the raw data for analysis after considering missing values and errors.

The NHIS-NHID provides patient residence information at the district level. Korea is divided into 17 regions according to the administrative system. We divided the spatial scope into seven cities (one special and six metropolitan cities). As of 2018, the seven cities accounted for approximately 5.4% (5,424 km$^2$) of the total area and approximately 43.9% (22.73 million) of the total population of Korea (19, 20). According to the regional classification of NHIS-NHID, the seven cities comprise 74 districts, with each city consisting of 5–25 districts.

The cause of CVD in this study is defined according to the International Classification of Diseases 10th codes (ICD-10). We analyzed the first occurrence of hospitalization for all cases of CVD (ICD-10: I00–I99) and stroke (ICD-10: I60–I64) in 2016–2018. In the NHIS-NHID, the disease of the patient is recorded as the principal diagnosis and there are one to four additional diagnoses; the principal is the disease that mostly affects a patient's symptoms, and the latter ones are optionally recorded diseases. In this study, we analyzed the patients whose principal and first additional diagnoses were recorded as CVD. Considering previous studies (21–23), we selected individual-level covariates such as age, sex, health insurance premiums, body mass index (BMI) (kg/m$^2$), total cholesterol (mg/dl), hypertension, diabetes, smoking status (non-smoker or smoker), and alcohol intake (non-drinker or drinker). For this, we used the medical checkup data of the NHIS-NHID. As NHIS provides medical checkups to those over 20 years old, this study only includes participants who are over 20 years old. In Korea, health insurance premiums are determined by household income; therefore, we considered health insurance premiums as the participant's economic level. For age, sex, and health insurance premiums, we used data from the year of disease occurrence or censoring (death or the end of the study). For the other covariates, we used the most recent available data among the medical checkup data in 2015–2018 because the NHIS provides the medical checkup service biennially, and since this medical checkup is not enforceable it is common for people to skip it. For covariates BMI and total cholesterol, we eliminated outliers by calculating upper and lower boundaries by taking three standard deviations from the mean (24).

For the analysis, we initially selected 2,288,781 people without a history of CVD in the period 2002–2015. Next, we excluded 43,689 people (1.9%), which are called "all participants" hereafter, due to a lack of PM2.5 measurement. Then, we excluded 759,884 people (33.2%) whose individual-level data, such as BMI, total cholesterol, etc., was missing. Finally, we obtained 1,485,208 participants (64.9%), which are called "selected participants" hereafter, for analysis. The disease-specific data selection process is shown in the **Supplementary Figure S1**.

PM2.5 concentrations in the participants' residences were estimated using the Ordinary Kriging (OK) method. The OK method was conducted from January 1, 2015 to December 31, 2017 based on hourly PM2.5 data of 242 monitoring stations in 2017. We used the hourly measured PM2.5 values provided by Korea Environment Corporation since 2015 and measured by beta-ray absorption methods. The OK method was conducted using the automap package in R (25). To evaluate the quality of the estimated PM2.5 values from the OK method, we used leave-one-out cross-validation, a technique that excludes one monitoring station at each step. Then, we estimated the PM2.5 concentrations of the excluded station using the measured values at the remaining stations and compared them with the measured value at the excluded station (**Supplementary Figure S2**).

The annual frequency of high PM2.5 concentration (FH) was calculated by the annual hours when the average PM2.5 concentration was above 55.5 μg/m$^3$ (hereafter, FH55). We assessed the PM2.5 exposure of each participant using the following: 1) the 1-year average PM2.5 exposure in 2015 (E1), 2) the 2-year average PM2.5 exposure for 2015–2016 (E2), and 3) the 3-year average PM2.5 exposure for 2015–2017 (E3). We considered E1 as our primary exposure period. We weighted the average PM2.5 exposure according to the time spent in a specific district if the participant moved their residence.

## Statistical Analyses

We estimated the hazard ratios (HRs) and 95% confidence intervals (CIs) using Cox proportional-hazards model developed by Cox (26). We used the weighted Schoenfeld residuals (27) to verify the proportional hazard assumption in each Cox model. We found that they did not violate the proportional hazard assumption, with all $p$-values exceeding 0.05. The survival time of each participant was calculated, in days, from January 1, 2016 to the earliest date of CVD occurrence or censoring (death or the end of the study on December 31, 2018).

We applied three analysis models according to the participants and individual-level covariates. In Model 1, we included only sex and age as covariates for all participants. In Model 2, we included only sex and age as covariates for selected participants, excluding some without individual-level data. In Model 3, we included covariates BMI, total cholesterol, hypertension, diabetes, smoking status, and alcohol intake, in addition to those in Model 2. We considered Model 3 as our primary model.

Moreover, we performed subgroup analysis by sex (Male or Female), age (Younger ≤ 65 years old; Older > 65 years old), BMI (Underweight < 18.5 kg/m$^2$; Normal 18.5–30 kg/m$^2$; Obese ≥ 30 kg/m$^2$), total cholesterol (Normal < 240 mg/dl; High ≥ 240 mg/dl), hypertension (No or Yes), diabetes (No or Yes), smoking status (non-smoker or smoker), and alcohol intake (non-drinker or drinker). All analyses were conducted using STATA (version 17; Stata Corporation, College Station, TX, United States).

## RESULTS

**Table 1** presents the characteristics of the study population, which includes 47,149 and 6,545 cases of all CVD and stroke, respectively. In the entire cohort without CVD, the average age of the study participants was 44.2 years, and 57.6% of them were men. Non-smokers constituted 58.1% of the participants, whereas smokers constituted 41.9%. In the female group, the proportion of smokers (6.2%) was minuscule compared with that of non-smokers (93.8%). Meanwhile, in the male group, the proportion of smokers (68.2%) was higher than that of non-smokers (31.8%). In addition, the proportion of hypertensive people in the cohorts with CVD was relatively higher than of those in the entire cohort without CVD. Interestingly, there were few drinkers in the cohorts with CVD.

**Table 2** presents the average PM2.5 and FH55 during the exposure period. The annual average PM2.5 varied according to the exposure periods; it was the highest for E2 (24.9 μg/m$^3$). Meanwhile, FH55 was the longest for E1 (274.8 h). To compare the effects of average PM2.5 and FH55, HRs were quantified for an interquartile range (IQR) change in each exposure. In addition, regardless of the exposure period, we used the IQR of E1 (2.9 μg/m$^3$ and 94 h).

**Table 3** presents the estimation results of the Cox model in terms of HRs and 95% CIs using three different models for E1. For all CVD, the IQR increase in average PM2.5 in Models 2 and 3 increased risk by 12.7% (95% CI, 10.8–14.6) and 11.6% (95% CI,

TABLE 1 | Characteristics of the study populations [National Health Insurance Service–National Health Information Database, South Korea, 2002–2018].

| Characteristic | Entire cohort without CVD (N = 1,431,514) | All CVD (I00–I99) (N = 47,149) | Stroke (I60–I64) (N = 6,545) |
|---|---|---|---|
| Sex [n (%)] | | | |
| Male | 824,238 (57.6) | 27,370 (58.1) | 4,409 (67.4) |
| Female | 607,276 (42.4) | 19,779 (41.9) | 2,136 (32.6) |
| Age [n (%)] | | | |
| Younger | 1,382,807 (96.6) | 40,713 (86.3) | 5,046 (77.1) |
| Older | 48,707 (3.4) | 6,436 (13.7) | 1,499 (22.9) |
| Average age of occurrence [years][a] | 44.24 (11.27) | 52.57 (12.19) | 56.96 (11.81) |
| Body mass index (BMI) [n (%)] | | | |
| Underweight | 67,669 (4.7) | 1,452 (3.1) | 219 (3.3) |
| Normal | 1,312,884 (91.7) | 43,361 (92) | 6,043 (92.3) |
| Obese | 48,552 (3.4) | 2,218 (4.7) | 227 (3.5) |
| Average BMI [kg/m$^3$][a] | 23.4 (3.24) | 24.05 (3.3) | 23.95 (3.21) |
| Total cholesterol [n (%)] | | | |
| Normal | 1,277,347 (89.2) | 40,030 (84.9) | 5,552 (84.8) |
| High | 154,167 (10.8) | 7,119 (15.1) | 993 (15.2) |
| Average total cholesterol [mg/dL][a] | 195.41 (34.31) | 200.8 (37.86) | 200.06 (39.54) |
| Hypertension [n (%)] | | | |
| No | 1,429,619 (99.9) | 45,134 (95.7) | 6,085 (93.0) |
| Yes | 1,895 (0.1) | 2,015 (4.3) | 460 (7.0) |
| Diabetes [n (%)] | | | |
| No | 1,416,878 (99.0) | 45,367 (96.2) | 6,157 (94.1) |
| Yes | 14,636 (1.0) | 1,782 (3.8) | 388 (5.9) |
| Smoking status [n (%)] | | | |
| Non-smoker | 831,701 (58.1) | 24,843 (52.7) | 2,922 (44.6) |
| Male non-smoker | 262,242 (31.8) | 6,496 (23.7) | 960 (21.8) |
| Female non-smoker | 569,359 (93.8) | 18,347 (92.8) | 1,962 (91.9) |
| Smoker | 599,813 (41.9) | 22,306 (47.3) | 3,623 (55.4) |
| Male smoker | 561,896 (68.2) | 20,874 (76.3) | 3,449 (78.2) |
| Female Smoker | 37,917 (6.2) | 1,432 (7.2) | 174 (8.1) |
| Alcohol intake [n (%)] | | | |
| Non-drinker | 576,068 (40.2) | 22,532 (47.8) | 3,139 (48.0) |
| Drinker | 855,446 (59.8) | 24,617 (52.2) | 3,406 (52.0) |

[a]Data is presented as mean (standard deviation).

TABLE 2 | Descriptive statistics for air pollutants during individual level exposure periods [National Health Insurance Service–National Health Information Database, South Korea, 2002–2018].

| Variables | Exposure period[a] | Mean ± SD | Interquartile range | Range |
|---|---|---|---|---|
| PM2.5 (μg/m$^3$) | E1 | 24.4 ± 1.6 | 2.9 | 21.9–27.6 |
|  | E2 | 24.9 ± 1.0 | 1.4 | 22.3–27.5 |
|  | E3 | 24.3 ± 1.1 | 1.1 | 16.7–27.0 |
| FH55 (hours) | E1 | 274.8 ± 95.6 | 94.0 | 164.0–565.0 |
|  | E2 | 271.6 ± 67.9 | 77.6 | 142.5–485.4 |
|  | E3 | 257.9 ± 52.1 | 49.1 | 138.0–427.3 |

Abbreviation: PM2.5, particulate matter less than 2.5 μm in diameter; FH55, the annual hours when average PM2.5 concentration was above 55.5 μg/m$^3$.
[a]The exposure periods are as follows: E1 for 2015; E2 for 2015–2016, E3 for 2015–2017.

9.7–13.6), respectively, which was higher than that for Model 1 (HR, 1.068; 95% CI, 1.056–1.080). Similarly, for FH55 exposure, the estimation results for selected participants (Models 2 and 3) were higher than those for all participants (Model 1). We found that for every IQR increase of FH55, the risk increased by 2.1% (95% CI, 1.5–2.7), 4.4% (95% CI, 3.5–5.3), and 3.8% (95% CI, 2.8–4.7) in Models 1, 2, and 3, respectively. These results were statistically significant at the 1% level. However, regarding stroke, most estimated results were not statistically significant.

Next, for our primary model, Model 3, we analyzed the effect of PM2.5 exposure over three different exposure periods. Table 4 presents the estimated HRs with 95% CIs. We found that, for all CVD, the shorter the exposure period, the higher the risk. We also found that, for every IQR increase in average PM2.5, the risk increased by 11.6% (95% CI, 9.7–13.6), 10.3% (95% CI, 6.8–13.9), and 7.9% (95% CI, 3.6–12.5) in E1, E2, and E3, respectively. Similarly, for FH55, the shorter the exposure period, the higher the HRs. However, regarding E3, the risk decreased as per the IQR

**TABLE 3** | Hazard ratios and 95% confidence intervals of cardiovascular disease for interquartile range increase in PM2.5 exposure in 2015 by models [National Health Insurance Service–National Health Information Database, South Korea, 2002–2018].

| Variable | Hazard ratios (95% confidence interval) | | |
| --- | --- | --- | --- |
| | Model 1[a] | Model 2[b] | Model 3[c] |
| Average PM2.5[d] | | | |
| All CVD (I00–I99) | 1.068 (1.056, 1.080)[e] | 1.127 (1.108, 1.146)[e] | 1.116 (1.097, 1.136)[e] |
| Stroke (I60–I64) | 1.009 (0.980, 1.038) | 1.050 (1.003, 1.099)[f] | 1.041 (0.995, 1.089) |
| FH55[d] | | | |
| All CVD (I00–I99) | 1.021 (1.015, 1.027)[e] | 1.044 (1.035, 1.053)[e] | 1.038 (1.028, 1.047)[e] |
| Stroke (I60–I64) | 1.001 (0.986, 1.017) | 1.016 (0.992, 1.041) | 1.009 (0.985, 1.034) |

Abbreviation: CVD, cardiovascular disease; FH55, the annual hours when average PM2.5 concentration was above 55.5 μg/m$^3$; CVD, cardiovascular disease.
[a]Model 1 included sex and age as covariates for all participants, including participants without individual-level data.
[b]Model 2 included sex and age as covariates for selected participants, excluding participants without individual-level data.
[c]Model 3 included sex, age, Body mass index, total cholesterol, hypertension, diabetes, smoking status, and alcohol intake as covariates for selected participants, excluding some participants without individual-level data.
[d]Interquartile ranges for PM2.5 and FH55 were 2.9 μg/m$^3$ and 94 h, respectively.
[e]Significant at 1% level.
[f]Significant at 5% level.

**TABLE 4** | Hazard ratios and 95% confidence intervals of cardiovascular disease for interquartile range increase in PM2.5 exposure by three exposure periods [National Health Insurance Service–National Health Information Database, South Korea, 2002–2018].

| Variable | Hazard ratios (95% confidence interval)[a] | | |
| --- | --- | --- | --- |
| | E1[b] | E2[b] | E3[b] |
| Average PM2.5[c] | | | |
| All CVD (I00–I99) | 1.116 (1.097, 1.136)[d] | 1.103 (1.068, 1.139)[d] | 1.079 (1.036, 1.125)[d] |
| Stroke (I60–I64) | 1.041 (0.995, 1.089) | 1.040 (0.954, 1.134) | 1.046 (0.937, 1.167) |
| FH55[c] | | | |
| All CVD (I00–I99) | 1.038 (1.028, 1.047)[d] | 1.033 (1.018, 1.048)[d] | 0.997 (0.971, 1.024)[d] |
| Stroke (I60–I64) | 1.009 (0.985, 1.034) | 1.002 (0.961, 1.043) | 1.032 (0.960, 1.109) |

Abbreviation: FH55, the annual hours when average PM2.5 concentration was above 55.5 μg/m$^3$; CVD, cardiovascular disease.
[a]Included sex, age, Body mass Index, total cholesterol, hypertension, diabetes, smoking status, and alcohol intake as covariates for selected participants, excluding participants without individual-level data.
[b]The exposure periods are as follows: E1 for 2015; E2 for 2015–2016, E3 for 2015–2017.
[c]Interquartile ranges for PM2.5 and FH55 were 2.9 μg/m$^3$ and 94 h, respectively.
[d]Significant at 1% level.

increase of FH55 (HR: 0.997, 95% CI: 0.971–1.024). These findings were statistically significant at the 1% level.

In addition, we conducted subgroup analysis using sex, age, BMI, total cholesterol, hypertension, diabetes, smoking status, and alcohol intake to identify those sensitive to average PM2.5 exposure. **Figure 2** presents the HRs and 95% CIs for all CVD by subgroups, and those of stroke are shown in the **Supplementary Figure S3**. We found that the risk of the female smoker group was the highest. In addition, the estimated HRs of the participants who were female, of older age, and smokers were higher than those of each comparative group. Most results were not statistically significant for stroke; however, we found that participants of older age and with a history of hypertension were significantly more vulnerable to average PM2.5 exposure.

## DISCUSSION

### Main Findings
In this study, we analyzed the effect of PM2.5 exposure on the first occurrence of hospitalization for CVD, considering average PM2.5 exposure and the frequency of high PM2.5 concentration using large-scale cohort data from seven metropolitan cities in Korea. We found that PM2.5 exposure, both average PM2.5 and the frequency of high PM2.5, increased the risk of all CVD occurrence. An IQR increase in average PM2.5 and FH55 was associated with an 11.6% and 3.8% increase in risk, respectively. We found that, regarding stroke, only a few groups were vulnerable to PM2.5 exposure.

From the analysis, we found that the first occurrence of all CVD is related to average PM2.5. The estimated risk in this study was higher than that in previous studies analyzing other countries. The reasons for this could be due to the following differences between this study and previous ones (7, 28–33). First, the average PM2.5 concentration in this study was higher than in studies conducted in western countries. In previous studies, the mean PM2.5 concentration exposures were 9.6 μg/m$^3$ in Canada (7); 13.4 μg/m$^3$ (28), 13.5 μg/m$^3$ (29), and 17.8 μg/m$^3$ (30) in the United States; and 10 μg/m$^3$ (31) and less than 19 μg/m$^3$ (32) in Europe except for Italy, whereas the mean PM2.5 concentration exposure in this study (Korea) was 22.4 μg/m$^3$. Second, the health endpoints and data coverage were different. In this study, we selected the first occurrence of hospitalization for all CVD and stroke as the endpoint. For example, in a recent study conducted in Hong Kong, Qiu et al. (33) selected the first occurrence of emergency hospital admissions as a health endpoint and used the health record data collected from public hospitals, excluding private hospitals. The NHIS-NHID data used in this study was more extensive, including those from not only public hospitals but also private hospitals. Third, Korea is one of the most densely populated countries among the Organization for Economic Cooperation and Development countries (34). In a recent study conducted in China, the cohort used included 15 provinces, including Beijing (1,324 people/km$^2$), which has a high population density and Qinghai (8.2 people/km$^2$), which is less densely populated (9, 35, 36). This study covered seven metropolitan cities in Korea that have a very high population density at more than 1,000 people/km$^2$ (37).

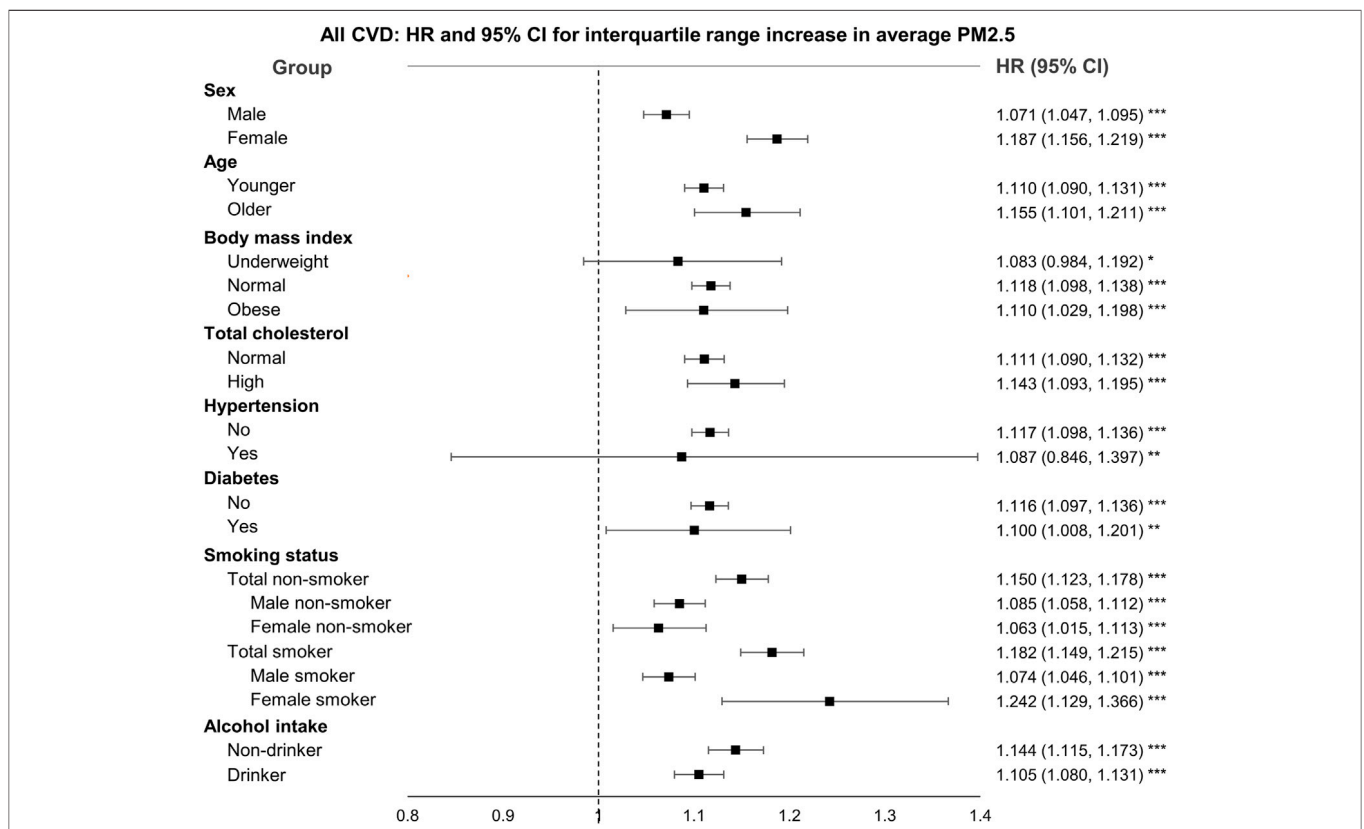

**FIGURE 2** | Hazard ratios and 95% confidence intervals for all cardiovascular disease by subgroups [National Health Insurance Service–National Health Information Database, South Korea, 2002–2018]. Abbreviation: CI, confidence interval; HR, hazard ratios; PM2.5, particulate matter less than 2.5 μm in diameter. Interquartile range was 2.9 μg/m$^3$. * Significant at 10% level. ** Significant at 5% level. *** Significant at 1% level.

The results of subgroup analysis are consistent with those of previous studies (7, 8, 29, 33, 38–40). In this study, the risk of the non-hypertensive group was higher than that of the hypertensive group. This result is similar to previous studies (7, 29, 38), but it contradicts other studies (8, 9, 39, 40). In terms of diabetes, we found that participants without diabetes were at a higher risk, similar to a previous study conducted in Seoul, Korea (8). However, a previous study of all Korean regions presented the opposite result (39). In terms of smoking status, the HR of the smoker group was higher than that of the non-smoker group; the risk of the female smoker group was the highest. These results contradict those of Qiu et al.

(33). They conducted the analysis, classifying the smoking status into never, former, or current smoker, and found that the risks of female former and current smoker groups were higher than those of male groups and female never smoker group. However, in the case of ischemic stroke, a subtype of stroke, the risk of the male current smoker group was higher than that of the female group. In contrast, some studies found that the HR of the non-smoker group was higher than that of the smoker group (8, 9, 40); regarding these results, Puett et al. argued that the health effects of smoking might dilute the health effects of air pollution (38).

Next, we also analyzed the effect of annual hours when the average PM2.5 concentrations was above 55.5 μg/m$^3$, which has not been examined in previous studies. We found that FH55 increased the risk of CVD occurrence, but had a relatively less effect compared with the average PM2.5. This finding could be attributed to people's perception and adaptive behavior to PM. Cho and Kim surveyed 171 Korean people using a five-point Likert scale (1 point: very unlikely–5 points: very likely) regarding their perception of PM and response behavior in Seoul, Korea (41). Respondents gave 4.00 and 3.43 points for "checking the PM concentration every day" and "refraining from going out when the PM concentration is high," respectively. Wells et al. analyzed 10,898 adults to determine if they tried to mitigate exposure when air pollution was severe using the data from the 2007–2010 National Health and Nutrition Examination Survey in the United States (42). They defined the susceptible groups as the older adults, people with respiratory disease, and people with CVD. They found that 14.2% of older adults, 25.1% of people with respiratory disease, and 15.5% of people with CVD changed their behavior according to atmospheric PM levels. In particular, 65.9% of those who said they changed their behavior reported that they "spent less time outdoors." Moreover, air quality alerts can help change the behavior of people when the air quality is very poor. Noonan's survey determined whether smog alerts could change people's behavior in Atlanta, United States (43). He found that the issuance of smog alerts reduced park usage in the older adults and exercising groups. Saberian et al. analyzed cycling use changes in Sydney due to air pollution (44). They analyzed the cycle movement count collected from electronic path-side devices installed at 31 points in the designated cycle paths from May 2008 to September 2013, and found a 14–35% reduction in cycle movement when air quality alerts were issued. Similarly, it appears that these perceptions and behaviors reduced the risk of disease occurrence despite the increased annual hours when the average PM2.5 concentration was above 55.5 μg/m$^3$. This means that people restrain outdoor activities and reduce PM exposure voluntarily when it is predicted that PM is high, and this may dilute the increased risk with an increase of FH55.

## Strengths and Limitations

This study has several strengths. First, to the best of our knowledge, it is the first study to analyze the effect of frequency of high air pollutant concentration. We found that the relationship between the frequency of high PM2.5 concentration and occurrence of all CVD was statistically significant, but had a relatively less effect compared to average PM2.5. It can be interpreted that people would reduce their risk by voluntarily avoiding high PM2.5 exposure by checking real-time PM2.5 concentration information and air quality alerts. Therefore, accurate forecasts, information dissemination, and timely warning of high PM2.5 concentrations at the national level have the potential to reduce the risk of CVD occurrence. Second, we used a large-scale cohort with 1,485,208 study participants. The NHIS is a universal coverage health insurance system that is mandatory for citizens in Korea. Most previous studies conducted in Korea used NHIS-NSC data, which includes one million randomly extracted people from the total population regardless of disease. However, this study used NHIS-NHID data, which includes all CVD patients from 2016 to 2018 without a previous history of the disease. With this large amount of data, we were able to estimate the effect of PM2.5 exposure on CVD more precisely and derive more reliable and generalizable results. Third, we considered the residential moves of the study population during exposure periods.

Despite the strengths, this study has a few limitations that could be overcome by further research. First, there have been no previous studies that analyzed the frequency of high air pollutant concentration; therefore, it is difficult to compare and generalize our results, which would be based on people's regional adaptive behaviors in Korea. The annual average PM2.5 concentration in Korea is decreasing gradually but is still higher than that in other countries. Likewise, the global annual average PM2.5 concentration is steadily decreasing (45, 46). However, cities in some countries still experience high PM2.5 concentrations for several hours of the year (13). In countries where PM2.5 measurement, prediction, and forecasting are not systematically performed, the frequency of high PM2.5 concentration may affect the occurrence of disease. Therefore, various follow-up studies are needed to investigate the relationship between the frequency of high air pollutant concentrations and human health. Second, we used PM2.5 concentrations in residential regions, and not workplace regions. As is the case for many workers, the region where they spend their most time working is different from their region of residence. This study considered only the residential region. Third, we did not include road traffic noise, which is known to affect CVD occurrence, as covariates in the model. Some previous studies found that exposure to road traffic noise affects the risk of CVD occurrence (31, 47–49). However, road traffic noise data in Korea is recorded in very few locations. Out of the 74 districts in the seven cities that we analyzed, automatic noise monitoring stations exist in only 28 of them. There are manual noise monitoring stations instead, which record six times a day (four times at daytime and two at nighttime) every quarter. We would have had to exclude more data to include road traffic noise as a covariate; therefore, we did not consider it. Lastly, we did not cover all participants because we excluded some whose individual-level data was unavailable. We compared the sex and age distributions of all the participants before exclusion and the selected participants after exclusion (**Supplementary Table S2**). Although the distribution was similar, the selected participants had a slightly higher proportion of men and those aged 30–60 years.

## Conclusion

As the annual frequency of high PM2.5 concentrations increases over time, it is crucial to confirm the relationship between the frequency of high PM2.5 concentrations exposure and human health. Using large-scale national cohort data, we focused on CVD occurrence and found that the annual average PM2.5 concentration affects CVD occurrence, and the frequency of high PM2.5 concentrations had less of an effect on CVD occurrence than the average PM2.5. This could be because people voluntarily avoided high PM concentrations exposure through real-time information and alerts.

# ETHICS STATEMENT

The studies involving human participants were reviewed and approved by the Institutional Review Board of Yonsei University. Written informed consent for participation was not required for this study in accordance with the national legislation and the institutional requirements.

# AUTHOR CONTRIBUTIONS

EC designed research, conducted statistics analysis, and wrote the manuscript. YK reviewed and edited the manuscript. YC supervised study, analysis, interpretation of the results, and provided critical revision of the manuscript. All authors read and approved the final version of manuscript.

## FUNDING

The authors declare that this study received no specific grant from any funding agency in the public, commercial, or not-for-profit sectors.

## CONFLICT OF INTEREST

The authors declare that the research was conducted in the absence of any commercial or financial relationships that could be construed as a potential conflict of interest.

## ACKNOWLEDGMENTS

This study used NHIS-NHID data (NHIS-2020-1-103) collected by the NHIS.

## SUPPLEMENTARY MATERIAL

The Supplementary Material for this article can be found online at: https://www.ssph-journal.org/articles/10.3389/ijph.2022.1604389/full#supplementary-material